\documentclass[fleqn,10pt]{wlscirep}
\usepackage{siunitx}
\usepackage{bm}
\newcommand{\latinphrase}[1]{\textit{#1}}  
\newcommand{\etal}{\mbox{\latinphrase{et al.\ }}}
\newcommand{\ie}{\mbox{\latinphrase{i.e.,\ }}}

\newcommand{\kb}{k_\textnormal{B}}
\newcommand{\ps}{P_\textnormal{s}}
\title{Heat production and error probability relation in Landauer reset at effective temperature}

\author[1,2,*,+]{Igor Neri}
\author[1,!,+]{Miquel L\'opez-Su\'arez}
\affil[1]{NiPS Laboratory, Dipartimento di Fisica e Geologia,
             Universit\`a degli Studi di Perugia, 
             06123 Perugia, Italy}
\affil[2]{INFN Sezione di Perugia, via Pascoli,
             06123 Perugia, Italy}

\affil[*]{igor.neri@nipslab.org}
\affil[!]{miquel.lopez@nipslab.org}

\affil[+]{these authors contributed equally to this work}

\begin{abstract}
The erasure of a classical bit of information is a dissipative process. The minimum heat produced during this operation has been theorized by Rolf Landauer in 1961 to be equal to $\kb T \ln 2$ and takes the name of Landauer limit, Landauer reset or Landauer principle. Despite its fundamental importance, the Landauer limit remained untested experimentally for more than fifty years until recently when it has been tested using colloidal particles and magnetic dots. Experimental measurements on different devices, like micro-mechanical systems or nano-electronic devices are still missing. Here we show the results obtained in performing the Landauer reset operation in a micro-mechanical system, operated at an effective temperature. The measured heat exchange is in accordance with the theory reaching values close to the expected limit. The data obtained for the heat production is then correlated to the probability of error in accomplishing the reset operation.
\end{abstract}
\begin{document}

\flushbottom
\maketitle

\thispagestyle{empty}

\section*{Introduction}
The minimum energy required to reset one bit of information represents one of the fundamental limits of computation arising when one bit of information is erased or destroyed. Starting with a system encoding two possible states with the same probability and finishing the procedure with only one possible state, the variation of entropy in the system is equal to the difference of entropy between the final state and the initial one, $\Delta S = \kb \ln 1 - \kb \ln 2 = -\kb \ln 2$. On average this reduction of entropy has to be accompanied with an increase of heat in the surrounding environment in order to not violate the second law of thermodynamics, $Q_\textnormal{L} \geq \kb T \ln 2$. This limit takes the name of Landauer limit and was theorized by Rolf Landauer in the 60's \cite{landauer1961irreversibility}, and remained untested for over fifty years. The development of computational methods allowed to evaluate tiny amounts of heat exchanged and recent technical advances in micro- and nano-fabrication made possible to recently test the validity of the Landauer principle. In particular, the first experimental verification of the Landauer principle has been carried out by B\'erut \etal using a colloidal particle trapped in optical tweezers \cite{berut2012experimental}. More recently Jun \etal presented similar results considering a colloidal particle in a feedback trap \cite{jun2014high}. Finally, the last experimental verification of the Landauer limit has been performed by Hong \etal in a nanomagnetic system \cite{Honge1501492}. In this case information is encoded in the magnetization state of the system while an external magnetic field is used in order to flip between two preferred magnetization states. The evaluation of the heat produced during the reset operation was demonstrated to be compatible with the Landauer limit. Even if there is no doubt nowadays on the validity of the Landauer principle, the test on micro-mechanical systems is still missing. We have recently shown that it is possible to use electro-mechanical devices to accomplish basic\cite{lopez2015operating} and complex logic operations\cite{lopez2015sub} with an arbitrarily low energy expenditure. Therefore the possibility to use this class of devices for memory storage completes the logic architecture for a complete computing device.

In the following we show the measurement on heat production when performing the reset operation in a novel memory unit based on a bistable mechanical cantilever at effective temperature. The results are in good agreement with the Landauer limit. The dependence of the dissipation with the error rate is also investigated showing a trend in accordance with the theory\cite{gammaitoni2015towards}.

\section*{Results}
\subsection*{Bistable micro-mechanical system}
\begin{figure}[ht]
\centering
\includegraphics[width=\linewidth]{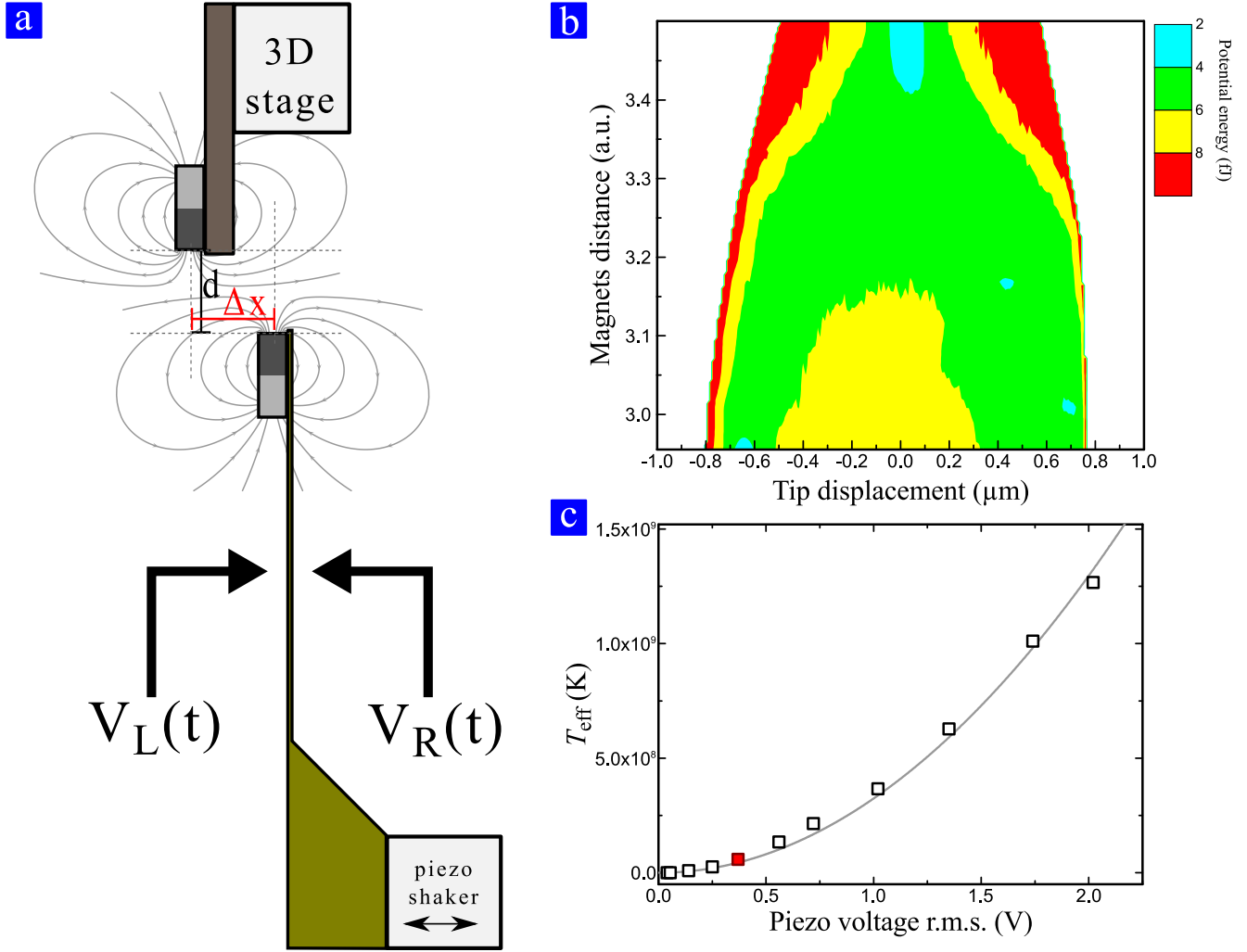}
\caption{Schematic of the whole system and measurement setup. Lateral view of the whole system and measurement setup. Two magnets with opposite magnetic orientations are used to induce bistability in the system. Two electrodes are used to apply electrostatic forces on the mechanical structure: $V_\textnormal{L}$ and $V_\textnormal{R}$ to force the cantilever to bend to the left (negative $x$) and to the right (positive $x$) respectively. The magnetic interaction can be engineered by changing geometric parameters such as $d$ and $\Delta x$. (b) Color-map of the reconstructed potential energy as function of the distance between the magnets, $d$. The distance is expressed in arbitrary units proportional to the voltage applied to the piezoelectric stage. Decreasing the distance between the magnets the potential energy softens and eventually two stable states appear. (c) Dependence of the effective temperature, $T_\textnormal{eff}$, with the root mean square of the white Gaussian voltage applied to the piezoelectric shaker. The red dot represent the condition accounted for the experimental data presented.}
\label{f:schematic}
\end{figure}

The mechanical system used to perform the experiment is depicted in Figure \ref{f:schematic} (a). A triangular micro-cantilever, \SI{200}{\micro\meter} long, is used to encode one bit on information. In order to obtain two stable states two magnets with opposite magnetization are placed on the tip of the cantilever and on a movable stage facing the cantilever. In this way, depending on the distance between the magnets, $d$, and the relative lateral alignment, $\Delta x$, it is possible to induce bistability on the system. Figure \ref{f:schematic} (b) shows the potential energy as a function of $d$ reconstructed from the probability density function of the position of the cantilever at equilibrium, $\rho (x,d) = A \exp(-U(x,d)/\kb T)$, which implies that $U(x,d)=-\kb T \ln \rho (x,d) + U_\textnormal{0}$\cite{berut2012experimental}. When the magnets are far away the effect of the repulsive force is negligible, the system is then monostable and can be approximated to a linear system. Decreasing the distance the repulsive force between magnets tends to soften the system up to the point where two stable positions appear. The effect of reducing even more $d$ is to enlarge the separation of the rest states and to increase the potential barrier separating these two wells. Eventually, when the distance between the magnets is small enough, the system remains trapped in one well for a period of time larger than the relaxation time of the system.
Logic states are encoded in the position of the cantilever tip: logic 0 for $x<0$ and logic 1 for $x>0$.
The proposed system presents intrinsic dissipative processes that depend on the maximum displacement of the cantilever tip\cite{lopez2015sub}. The minimum heat produced when performing a physical transformation of the system is proportional to $x_\textnormal{max}^2$. In our setup it is not possible to reduce the separation between the two potential wells to a value that bounds the heat produced by intrinsic dissipation below the Landauer limit. Increasing the effective temperature increases the value for the Landauer limit making possible to have negligible intrinsic dissipation. A piezoelectric shaker is used to excite the structure with a band limited white Gaussian noise to mimic the effect of an arbitrary temperature. In the present experiment the white noise is limited to \SI{50}{\kilo\Hz}, well above the resonance frequency of the free cantilever ($f_\textnormal{0}$=\SI{5.3}{\kilo\Hz}). The dependence of the effective temperature with the root-mean-squared voltage supplied to the shaker is reported in Figure \ref{f:schematic} (c). The red dot, corresponding to an effective temperature of $T_\textnormal{eff}$ = \SI{5e7}{\kelvin}, highlights the condition considered in the present case. The solid line represent the expected trend where $T \propto V_\textnormal{rms}^2$ \cite{venstra2013stochastic}. The effective temperature has been estimated computing the power spectral density (PSD) of the system at various piezoelectric noise excitation voltages. The obtained curves have been fitted with Lorenzian curves taking as reference for the calibration the one at room temperature ($T$=\SI{300}{\kelvin}). The other curves have been used to extract the only varying parameter $T_\textnormal{eff}$, corresponding to the effective temperature of the system under external excitation.
Finally two electrostatic probes, placed one on the left and the other on the right of the cantilever, are used to apply a negative and positive forces respectively. When a voltage different to zero is applied on one probe the cantilever feels an attractive electrostatic force toward the probe due to the polarization of the cantilever itself. The voltage on the probes, the distance between the magnets and their time evolution are used to specify the protocols used in order to change the bit stored in the system as described in the following subsection.

\subsection*{Reset protocol}
\begin{figure}[ht]
\centering
\includegraphics[width=0.5\linewidth]{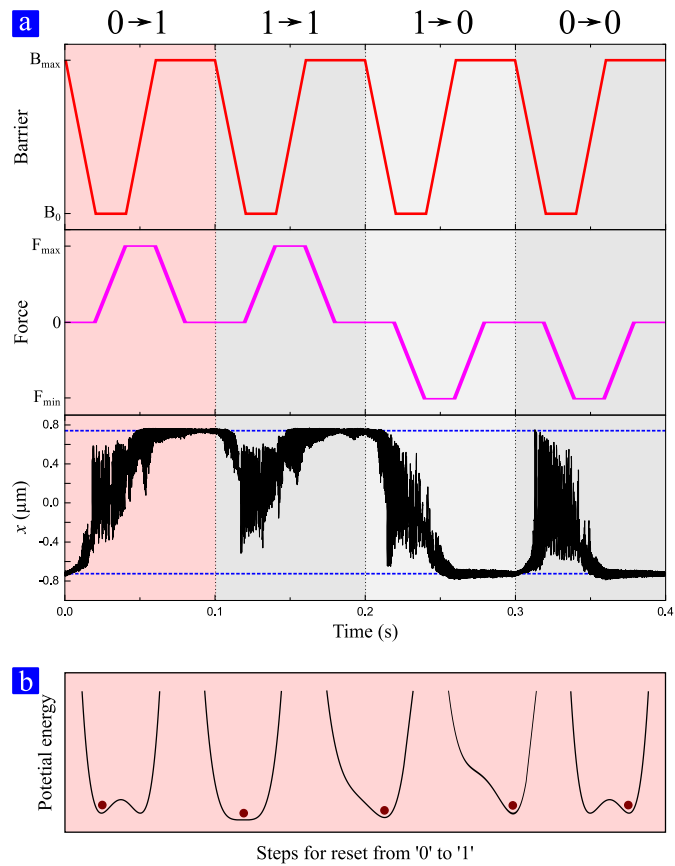}
\caption{Reset protocol. (a) Protocol used to perform the reset operation. In order to account for all the possible transitions we considered the reset to 1 (first two columns) and reset to 0 (last two columnss) starting from both 0 and 1 states. The first row depict the protocol used for removing the barrier. Once the barrier is removed a lateral force is applied (second row). The resulting displacement of the cantilever tip is represented in the third row. Once all the forces are removed and the barrier is restored the cantilever tip encodes the final state. (b) Schematic time evolution of the potential energy and state of the system for the case presented in the first column of panel (a). The operation starts from a double well potential and in the first step of the protocol the barrier is removed. During the next step the potential is tilted and the barrier is restored to its initial value. Finally, the lateral force is removed recovering the initial condition where the barrier between wells is at its maximum and the electrostatic forces are equal to zero.}
\label{f:protocol}
\end{figure}

Varying the magnets distance $d$ the barrier separating the two stable wells varies from the minimum value $B_\textnormal{0}$ for $d$=3.65 a.u. to $B_\textnormal{max}$ for $d$=2.8 a.u. Applying a voltage on one probe corresponds to apply a force toward the probe itself. Thus a voltage on the left probe corresponds to a negative force and a voltage on the right probe corresponds to a positive force. Details of the force calibration are presented in methods section.
In Figure \ref{f:protocol} (a) the protocol followed to reset the bit to the state 0 and 1 is presented. The procedure is similar to the ones presented in Refs. \citen{bennett1982thermodynamics} and \citen{berut2012experimental}. Initially the barrier separating the two stable states is removed moving the magnet away (red curve in the first panel) making the system monostable. Once the barrier is removed we apply a negative (positive) force to reset the bit status to 0 (1) applying a finite voltage $V_\textnormal{L}$ ($V_\textnormal{R}$) on the left (right) probe. This is represented by the magenta curve in Figure \ref{f:protocol} (a). Once the force is applied we restore the barrier to its original value. Finally, we remove the lateral force finishing in the original parameters configuration. At the end of the operation, if there are no errors the cantilever position encodes the desired bit of information. In order to be sure to perform the operation starting from both initial states, we mimic a statistical ensemble where the initial probability is 50\% to start in the left well and 50\% in the right well. A trace of the cantilever tip position, $x$, is shown in Figure \ref{f:protocol} (a) (black line), where the dashed blue lines represent the two stable positions once the barrier is restored. When the barrier is removed the system goes from the local prepared state to an undefined state with a free expansion, the entropy on the system thus increases in a irreversible manner\cite{bennett1982thermodynamics,crooks1999}. This increment is related to uncontrollable transitions from one well to the other once the barrier height is close to $\kb T$. These large excursions of the cantilever can be seen in the time series of position. Figure \ref{f:protocol} (b) shows a representation of the time evolution of the potential energy during the set of the bit to the logic state 1. In a first step the barrier is removed allowing the system to oscillate in a monostable potential landscape. Then the potential is slightly tiled and when the barrier is restored the system is confined in the desired state. After these stages the bit is set to the state 1. In order to have a reliable measure of the heat produced during the considered operations, reset protocols are repeated for 800 times in order to have a large enough statistic.

\subsection*{Heat vs error probability}
In the optimal case the initial configuration is a mixed logical 0 and 1 where both states have the same probability while the final configuration is the selected state with a $100\%$ probability. This corresponds to an entropy variation of $\Delta S = -\kb \ln(2)$ and a minimum heat produced of $Q\ge -T\Delta S = Q_\textnormal{L}$. It has been shown that the bound  $Q_\textnormal{L} = -\kb T \ln(2)$ applies only for symmetric potentials. Considering asymmetries on the system the minimum produced heat can be lowered below $Q_\textnormal{L}$ \cite{sagawa}. In our setup the system is slightly asymmetric and we have evaluated the variation of entropy from the initial to the final state from the probability density function of the tip position, $\rho(x)$, being $\Delta S_\textnormal{G} = -0.61\kb $ $\Delta S_\textnormal{S} = -0.68\kb$ for the Gibbs and Shannon entropy respectively, both close to $-\kb \ln(2)$.

If we consider the possibility to commit errors during the reset operation the heat produced becomes a function of the probability of success\cite{gammaitoni2015towards}:
\begin{equation}
Q(\ps)\ge \kb T \left[ \ln(2) + \ps \ln (\ps) + (1-\ps) \ln (1-\ps) \right]
\label{e:Q_ps}
\end{equation}
where $\ps$ is the probability of success or success rate. When $\ps$ is 0.5 no reset operation is performed and thus there is no minimum heat to be produced during the operation \cite{jun2014high,gammaitoni2015towards}.

\begin{figure}[ht]
\centering
\includegraphics[width=\linewidth]{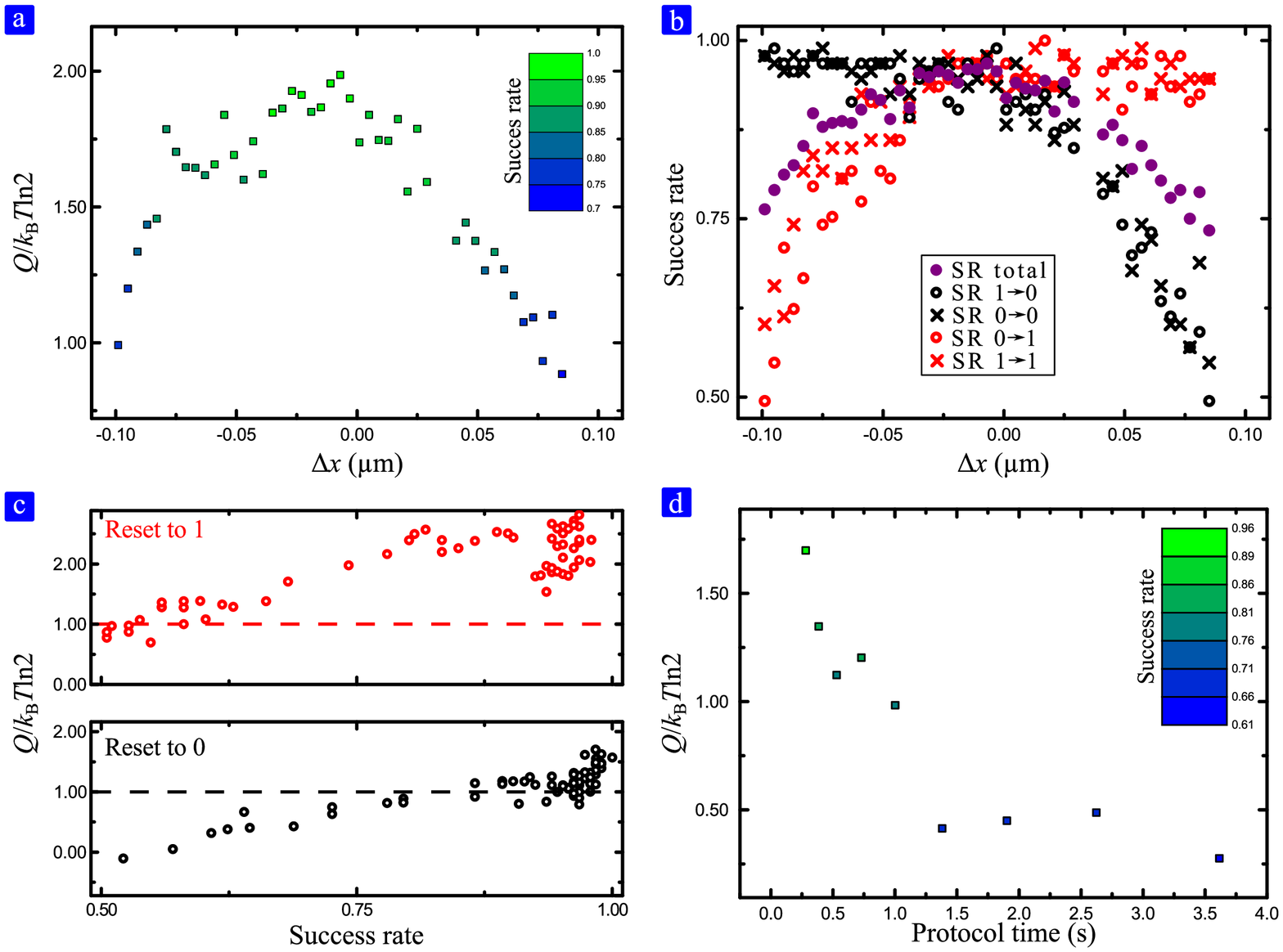}
\caption{Produced heat and probability of success for the reset operation. (a) Average heat produced during the reset operation as function of the lateral alignment $\Delta x$. For $\Delta x < 0$ the counter magnet is moved to the right and the 0 state ($x<0$) is favorable. Accordingly, for $\Delta x > 0$ the 1 state is more favorable. Introducing an asymmetry on the potential $Q$ decreases, which is accounted to the probability of success, $\ps$, that tends to decrease ($\ps$ is encoded in the color map). (b) Success rate of the reset operation as function of lateral alignment. Solid violet circles represent the overall success rate while black and red symbols account for the success rate resetting to 0 and 1 state respectively. The maximum overall success rate is present when the system is almost symmetric, $\Delta x \approx 0$. (c) Relation between success rate and heat dissipated. Red circles correspond to the resetting to 1 case while black ones correspond to the resetting to 0. (d) Dependence of $Q$ with the protocol time duration, $\tau_\textnormal{p}$. As $\tau_\textnormal{p}$ is increased the effects of frictional phenomena becomes negligible and the produced heat should approach the thermodynamic limit. However, for large $\tau_\textnormal{p}$ the reset operation fails giving a wrong logic output. In these cases, where the error probability is high, the produced heat is clearly below the Landauer limit. Inset shows the obtained relation between error probability (1-$P_\textnormal{s}$) and produced heat. The data are compatible with the minimum energy required for a given error probability as predicted by Eq. \ref{e:Q_ps}, represented by dashed line.}
\label{f:Q_vs_sr}
\end{figure}

In Figure \ref{f:Q_vs_sr} (a) we present the average heat produced for the reset operation as function of the lateral alignment of the counter magnet $\Delta x$. When the system is aligned closely to perfection (\ie $\Delta x \approx 0$) we estimate a heat production slightly above $\kb T$ and below two times $Q_\textnormal{L}$. Asymmetrizing the potential, by means of setting $\Delta x \neq 0$, the heat produced tends to decrease reaching values this time below $Q_\textnormal{L}$. However, in this conditions the error rate in performing the reset operation have a major role, in fact in this configuration the probability of success, $\ps$, decrease rapidly. This is represented by the color map of dots in Figure \ref{f:Q_vs_sr} (a), where green represents higher success rate while blue represents a higher probability of error. In Figure \ref{f:Q_vs_sr} (b) the success rate of the reset operation is reported as function of the lateral alignment. Solid violet circles represent the overall success rate while red and black symbols represent the error rate for resetting to 1 or to 0 respectively. Circles are used to report the error probability for the same initial and final state while crosses are used for 0 to 1 and 1 to 0 transitions. For instance let us consider the case where $\Delta x <0$: the counter magnet is moved towards the right and as a consequence the 0 state is more favorable respect the 1 state. From Figure \ref{f:Q_vs_sr} (b) we can see that for $\Delta x <0$ the probability of resetting toward 0 is almost 100\% while the probability of resetting toward 1 decreases rapidly reaching values below 50\%. The same behavior is present in the case $\Delta x > 0$, where the counter magnet is moved to the left, where the state 1 is more favorable.

We can now correlate the heat produced to the probability of success for resetting to 0 and 1 as presented in Figure \ref{f:Q_vs_sr} (c). Dashed lines represent the Landauer limit for a 100\% of success rate ($\approx 0.7 \kb T$). While in both cases the heat produced is above the Landauer limit, in the reset to 0 case the obtained values are very close to $Q_\textnormal{L}$. As expected, decreasing the success rate the obtained values goes below the Landauer limit for both cases accordingly to Equation \ref{e:Q_ps}.\\
As it is well known the adiabatic limit in presence of dissipation mechanisms like viscous damping can be only reached if the operation is performed slowly when these mechanisms are negligible. We increased the protocol time for the reset operation from \SI{0.25}{\second} up to \SI{3.5}{\second}. The results are presented in Figure \ref{f:Q_vs_sr} (d). Increasing the protocol time decreases the heat production reaching values well below the Landauer limit. Notice that in these cases where $Q < Q_\textnormal{L}$ the $\ps$ is well below 1 since the system has more time to relax and therefore tends to thermalize before the reset operation is correctly performed. In fact for protocols lasting more than \SI{1}{\second} the success rate is below 75\%.
\section*{Discussion}
We have measured the intrinsic minimum energy dissipation during the reset of one bit of information in a micro-mechanical system. We have considered a completely different physical system respect to the existing literature, \ie micro-electro-mechanical system. To achieve these results we have performed the experiment at an effective temperature of \SI{5e7}{\kelvin} in order to make the intrinsic dissipation of the mechanical structure negligible respect to the thermodynamic contribution. In these conditions we have reached values of heat produces consistent with the Landauer limit approaching it closely. Moreover we presented experimental data relating the minimum heat produced with the probability of success of resetting one bit of information. Nowadays where there is a lot of attention on micro-electro-mechanical systems able to perform computation at arbitrary low energy, the achieved results have a significant importance in the development of new computing paradigms based on systems different from the well established CMOS technology.

\section*{Methods}
\subsection*{Setup preparation and calibration}
The micro-cantilever used is a commercial atomic force microscopy (AFM) probe (NanoWorld PNP-TR-TL\cite{nanoworld}). It is long \SI{200}{\micro\meter}, with a nominal stiffness $k$=\SI{0.08}{\newton\per\meter} and a nominal resonace frequency of \SI{17}{\kilo\Hz}. A fragment of NdFeB (neodymium) magnet is attached to the cantilever tip with bi-component epoxy resin. To set the magnetization to a known direction the system is heated up to \SI{670}{\kelvin}, above its Curie temperature \cite{ma2002recent} in the presence of a strong external magnetic field with the desired orientation. With this additional mass the resonance frequency decreases to \SI{5.3}{\kilo\Hz}. The quality factor of the system has been estimated from the power spectral density of the displacement, $x$, fitted with a Lorenzian curve, giving a value of $Q_\textnormal{f}=320$.
The deflection of the cantilever, $x$, is determined by means of an AFM-like laser optical lever. A small bend of the cantilever provokes the deflection of a laser beam incident to the cantilever tip that can be detected with a two quadrants photo detector. The laser beam is focused on the cantilever tip with an optical lens (focal length $f=$\SI{50}{\milli\meter}). For small deflections the response of the photo detector remains linear, thus $x=r_\textnormal{x}\Delta V_\textnormal{PD}$, where $V_\textnormal{PD}$ is the voltage difference generated by the two quadrants of the photo detector. In order to determine $r_\textnormal{x}$ we look at the frequency response of the system as in \cite{lopez2015sub} under thermal excitation. The relation between the measured voltage and the expected displacement gives $r_\textnormal{x}=$\SI{1.8365e-5}{\meter\per\volt}.
The system is placed in a vacuum chamber  and isolated from seismic vibrations to maximize the signal-to-noise ratio. All measurements were performed at pressure $P=$\SI{4.7e-2}{\milli\bar}.
\subsection*{Effective temperature estimation}
As the system is modeled as a harmonic oscillator with one degree of freedom, according to the Equipartition Theorem the thermal energy present in the system is simply related to the cantilever fluctuations as $ \frac{1}{2}\kb T = \frac{1}{2} k \langle x^2\rangle $. According to Parseval's theorem $\langle x^2\rangle =\int_{0}^{\infty}\! \vert G(\omega) \vert ^2 \, \mathrm{d}\omega$ where $G(\omega)$ stands for the PSD of the system. This takes the form of a Lorenzian function $$ \vert G(\omega) \vert=  \sqrt{\frac{4\kb T}{Q_\textnormal{f} k \omega_\textnormal{0}}} \sqrt{\frac{1}{\left(1-\frac{\omega^2}{\omega_\textnormal{0}^2}\right)^2 + \frac{1}{Q_\textnormal{f}^2}\frac{\omega^2}{\omega_{0}^2}}} $$ where $\omega = 2 \pi f$.
Once the system has been calibrated at room temperature we estimated the effective temperatures fitting the only free parameter, $T$, from the measured PSDs to the expected function $\vert G(\omega) \vert$.
\subsection*{Force calibration}
A set of two electrodes (see Fig.~\ref{f:schematic}(a)) is used in order to polarize the cantilever producing a bend on the mechanical structure. Electrostatic forces depend on the voltage applied to the electrodes, i.e.\ $V_\textnormal{L}$ and $V_\textnormal{R}$ for the left and right electrode respectively. Since the restoring force of the cantilever can be expressed as $F_\textnormal{k}=-k x$, the relation between applied voltage to the probe and the force acting on the cantilever has been estimated in static conditions assuming $F_\textnormal{k}=F_\textnormal{el}$. The relation between the applied voltages, $V_\textnormal{L}$ and $V_\textnormal{R}$, and the electrostatic force $F_\textnormal{el}$ is then fitted with a $9^\textnormal{th}$ degree polynomial.
\subsection*{Heat production evaluation}
The work performed on the system along a given trajectory $x(t)$ is given by the integral\cite{seifert2012stochastic,douarche2005experimental}:
\begin{equation}
W=\int_{0}^{\tau_\textnormal{p}}{\sum_{k=1}^{M} \frac{\partial U(x,\boldsymbol{\lambda})}{\partial \lambda_k} \frac{\partial \lambda_k}{\partial t} \textnormal{d}t}
\end{equation}
where $\tau_\textnormal{p}$ is the protocol time duration, $U(x,\boldsymbol{\lambda})$ is the total potential energy of the system and $\boldsymbol{\lambda}$ is a vector containing all the $M$ control parameters. In our case we have two controls parameter, the voltage applied to the piezoelectric stage to control the energy barrier, and the electrostatic forces. To obtain the heat produced $Q$ we have to consider the variation in internal energy, $\Delta E$. Ideally the potential energy at both the bottom wells is the same, however considering asymmetries on the system $\Delta E$ can be different from zero. The evaluation of the total energy variation $\Delta E$ is obtained from the reconstructed potential energy, $U$, and the variation of the kinetic energy. The latter quantity is however negligible even for the shortest $\tau_p$, where the total kinetic energy variation is one order of magnitude lower than  $Q_\textnormal{L}$ (\SI{1.4e-17}{\joule} versus \SI{5.3e-16}{\joule}). Finally the heat produced is obtained from $Q=W-\Delta E$.

\section*{Acknowledgements}
The authors gratefully acknowledge useful discussion with L. Gammaitoni. The authors acknowledge financial support from the European Commission (FPVII, Grant agreement no: 318287, LANDAUER and Grant agreement no: 611004, ICT- Energy).

\section*{Author contributions statement}
M.L.S. and I.N. designed the experiment, performed the measurements, analyzed the measured data and contributed to the writing of the manuscript.

\section*{Data availability statement}
The data that support the findings of this study are available from the corresponding author upon request.

\section*{Additional information}
\textbf{Competing financial interests} The authors declare no competing financial interests.

\begin{thebibliography}{10}
%

\bibitem{landauer1961irreversibility} Landauer, R. {Irreversibility and heat generation in the computing process}. \emph{IBM journal of research and development} \textbf{5}, 183--191{1961}.

\bibitem{berut2012experimental} {B{\'e}rut, A.} \emph{et~al.} {Experimental verification of Landauer's principle linking information and thermodynamics}. \emph{{Nature}} \textbf{{483}}, {187--189} ({2012}).

\bibitem{jun2014high} {Jun, Y.}, {Gavrilov, M.} \& {Bechhoefer, J.} {High-precision test of Landauer’s principle in a feedback trap}. \emph{{Physical review letters}}  \textbf{{113}}, {190601}  ({2014}).

\bibitem{Honge1501492} {Hong, J.}, {Lambson, B.}, {Dhuey, S.} \& {Bokor, J.} {Experimental test of Landauer's principle in single-bit operations on nanomagnetic memory bits}. \emph{{Science Advances}} \textbf{{2}} ({2016}).

\bibitem{lopez2015operating}{ Lopez-Suarez, M.}, {Neri, I.} \& {Gammaitoni, L.} {Operating micromechanical logic gates below k{B}T: Physical vs logical reversibility}. In \emph{{Energy Efficient Electronic Systems (E3S), 2015 Fourth Berkeley Symposium on}}, {1--2}  ({IEEE}, {2015}).

\bibitem{lopez2015sub} {Lopez-Suarez, M.}, {Neri, I.} \& {Gammaitoni, L.} {Sub k{B}T micro electromechanical irreversible logic gate}. \emph{Nature Communication} \textbf{7} {12068} ({2016}).

\bibitem{venstra2013stochastic} {Venstra, W.~J.}, {Westra, H.~J.} \& {van~der Zant, H.~S.} {Stochastic switching of cantilever motion}. \emph{{Nature communications}} \textbf{{4}} ({2013}).

\bibitem{bennett1982thermodynamics} {Bennett, C.~H.} {The thermodynamics of computation—a review}. \emph{{International Journal of Theoretical Physics}} \textbf{{21}}, {905--940} ({1982}).

\bibitem{crooks1999} Crooks, Gavin E. {Entropy production fluctuation theorem and the nonequilibrium work relation for free energy differences}. \emph{Physical Review E} \textbf{60.3} {2721} ({1999}).

\bibitem{gammaitoni2015towards} {Gammaitoni, L.}, {Chiuchi{\'u}, D.}, {Madami, M.} \& {Carlotti, G.} {Towards zero-power ICT}. \emph{{Nanotechnology}} \textbf{{26}}, {222001} ({2015}).

\bibitem{sagawa} Sagawa, Takahiro. Thermodynamic and logical reversibilities revisited. \emph{Journal of Statistical Mechanics: Theory and Experiment} \textbf{2014.3} P03025 {(2014)}

\bibitem{nanoworld} {{NanoWorld - AFM tip - PNP-TR-TL - Pyrex-Nitride}}. {http://www.nanoworld.com/pyrex-nitride-triangular-silicon-nitride-tipless-cantilever-afm-tip-pnp-tr-tl}.
{Accessed: 2016-05-30}.

\bibitem{ma2002recent} {Ma, B.} \emph{et~al.} {{Recent development in bonded NdFeB magnets}}. \emph{Journal of magnetism and magnetic materials} \textbf{239}, {418--423} ({2002}).

\bibitem{seifert2012stochastic} {Seifert, U.} {Stochastic thermodynamics, fluctuation theorems and molecular machines}. \emph{{Reports on Progress in Physics}} \textbf{{75}}, {126001} ({2012}).

\bibitem{douarche2005experimental} {Douarche, F.}, {Ciliberto, S.}, {Petrosyan, A.} \& {Rabbiosi, I.} {An experimental test of the Jarzynski equality in a mechanical experiment}. \emph{{EPL (Europhysics Letters)}} \textbf{{70}}, {593} ({2005}).

\end{thebibliography}
\end{document}